\documentclass[twocolumn,prb,showpacs]{revtex4}
\usepackage{graphicx}
\usepackage{epsfig}
\usepackage{dcolumn}
\usepackage{bm}

\begin{document} 
\title{Phonon Transport in Amorphous-Coated Nanowires: \\an Atomistic Green Function Approach}
\author{N. Mingo and Liu Yang}
\affiliation{Eloret Corp., Mail Stop N229-1, NASA-Ames Research Center, Moffett Field, CA 94035-1000}
\begin{abstract}                
An approach is presented for the atomistic study of phonon transport in real
dielectric nanowires via Green functions. The formalism is applied to investigate the
phonon flow through nanowires coated by an amorphous material.
Examples for a simple model system, and for real Si nanowires coated by silica are given.
New physical results emerge for these systems, regarding the character
of the transition from ballistic to diffusive transport, the low temperature thermal
conductance, and the influence of the wire-coating interface on the thermal transport.
An efficient treatment of phonon scattering by the amorphous coating is also
developed, representing a valuable tool for the investigation of thermal conduction through
amorphous-coated nanowires.
\end{abstract}


\maketitle

\section{\label{sec:Intro}Introduction}

The problem of phonon transport through dielectric wires of nanometer thickness \cite{Lieber} 
is of special interest at present,
since accurate measurements of thermal conduction in these systems begin to be available \cite{Schwab,Fon,Kim}.
Theoretically, this problem has been considered by different approaches.
Some of these are the Boltzmann transport equation (BTE) \cite{Balandin}, Molecular Dynamics (MD) \cite{Volz,Schelling}, 
and the transmission function approach \cite{Kambili,Santamore,Rego,Ciraci,Angelescu,Blencowe}. The BTE has been succesfully used
in nanowire transport at high temperatures, for systems where the resistive length of the wire is long enough for the transport
to be diffusive. Molecular Dynamics has the advantage that it can accurately consider the anharmonic interactions between the atoms.
MD was applied to nanowires in ref.~\onlinecite{Volz}, for example. It is nonetheless difficult to study low temperatures using MD,
since it provides a classical description of the system.

The transmission function approach is very well suited to study cases when phonons flow ballistically or semi-ballistically.
In the limit of very low temperatures, an elastic continuum model provides a good description of the phonon flow.
This model has yielded new insight into the problem of phonon scattering by surface roughness, for example, where the 
transition from ballistic to diffusive transport \cite{Kambili} and the low temperature "dip" in the quantized
thermal conductance \cite{Santamore} have been investigated. Transmission function approaches have
also been succesfully applied to study
the heat flow through a mesoscopic link \cite{Patton}, a nanocristal \cite{Leitner}, and monatomic chains \cite{Ciraci}.
Despite these excellent works, and unlike the case of electron transport,
the use of the transmission function approach in phonon transport is still scarce.

In this paper we develop a novel transmission function approach, and apply it to 
study a problem that has not been theoretically investigated before:
the phonon transport along nanowires in which
part of the length is surrounded by a thick coating of amorphous material.
This is an important problem,
since most dielectric nanowires are naturally coated by a layer of amorphous material \cite{Lieber,Ma}.

After explaining the general formalism, its use is illustrated by the study of phonon transmission through 
two concrete systems: a 1-D chain partially coated with
an amorphous layer, and real 3-D nanowires with part of their length coated by a silica layer.
Three main phenomena are investigated:

-the transition from ballistic to diffussive transport. Explicit curves are shown of how this transition takes place,
as a function of frequency and coated length.

-the low temperature conductance. It is found that a phenomenon similar to the "dip" occuring in rough-edge 
uncoated wires \cite{Santamore}, occurs also in amorphous coated wires, although displaying some differences with
respect to the former case.

-the effect of the wire-coating interface. We find an interesting saturation effect as a function of the
coupling strength between the atoms at the interface. We also show the important influence of the interatomic link
number and configuration on the phonon transmission.

Our approach is also new in many respects:

1) An atomistic description is used. This is necessary since we are interested in how the
atomic structure of the interface between the wire and the coating affects transport. This also enables
to consider the whole dispersive spectrum, and not only the lower frequencies. Although excellent atomistic
investigation of transport through monatomic chains was done in ref.~\onlinecite{Ciraci}, we have not been able
to find any other transmission function atomistic study of larger systems, like the Si nanowires considered here. 
Despite the method is similar to that of tight binding electron transport, it is important to provide a 
complete and independent derivation of the formulas in the phonon framework, specially for those readers
not familiar with the theory of electron transport. Part of the paper is devoted to 
do this in detail.

2) We present a new method that allows to treat the amorphous overlayer in an efficient manner. As we will see,
the presence of the amorphous overlayer introduces a problem of overwhelming computational demand. We develop a technique 
that is able to yield the thermal conductance in the limit of thick overlayer, with much less computational effort.
The validity of this approach is explicitly shown with one example.

The structure of the paper is as follows.
Section~\ref{sec:method} explains the formalism. An introduction to the transmission function concept 
(Sect.~\ref{sec:trans}) is followed
by the core of the method, Section~\ref{sec:calc}, where we derive the formulas to compute
the phonon transmission function. 
Afterwards, Section~\ref{sec:SCflux}
describes the specific self-consistent technique that enables us to treat the amorphous coating surrounding
the wires. Section~\ref{sec:results} shows results for the two examples treated.
Conclusions are summarized in Section~\ref{sec:conclusions}.

\section{\label{sec:method}Method}

\subsection{\label{sec:trans}The transmission function in thermal conductivity and conductance}

Both the thermal conductivity and the thermal conductance can be obtained by the
same single approach. Let us imagine a perfectly harmonic and translationally invariant nanowire,
totally free of defects. The wire is free standing, so that there
is no heat leakage from the boundary. The harmonicity assumption implies that, unless there is
scattering due to disordered defects or impurities, the thermal conductance of the whole wire would be
independent of its length. In this situation, the heat flux associated with a small temperature
difference $\Delta T$ between the two wire ends is given by the sum of the contributions
of the individual phononic subbands \cite{Rego}:

\begin{eqnarray}
&&J_Q = \nonumber\\
&&\sum_\alpha \int_0^{\pi/a} 
\Delta T{\partial\over\partial T}\left [{1 \over {e^{\hbar \omega_\alpha(k) / k_B T} - 1}} \right ]
\nonumber \times {\partial\omega_\alpha\over\partial k}\hbar
\omega_\alpha (k) {dk\over 2\pi} \\
\nonumber &&  = \Delta T\sum_\alpha \int_{\omega_\alpha^{init}}^{\omega_\alpha^{fin}}
{\partial\over\partial T}\left [{1 \over {e^{\hbar \omega_\alpha(k) / k_B T} - 1}}\right ]\times
\hbar\omega {d\omega\over 2\pi},
\end{eqnarray}
where $\alpha$ labels the phonon subbands, $a$ is the wire's unit cell length, 
$k_B$ is Boltzmann's constant, $k$ is the wave vector in the
axial direction, $\omega_{\alpha}(k)$ is the dispersion relation of subband $\alpha$, 
$\omega_{\alpha}^{init(fin)}$ is the lower (upper) frequency
limit of the subband,
and $\omega$ is the frequency.
We note that the phonons inside the wire have a non equilibrium distribution different
from the the Bose-Einstein type. As it is clear from the second line, it is not necessary to know
the dependence of $\omega_\alpha$ on $k$, but only the
frequency limits of the subband.

The (length independent) thermal conductance for this limit is then
\begin{eqnarray}
\sigma_{sat} = \sum_\alpha \int_{\omega_\alpha^{init}}^{\omega_\alpha^{fin}} {\partial\over\partial
 T}
\left [{1 \over {e^{\hbar \omega / k_B T} - 1}} \right ]  \hbar\omega {d\omega\over 2\pi}.
\label{eq:sigmasat}
\end{eqnarray}

In general, however, one has a nanowire with disordered scattering sources along a finite length.
In such a case we no longer have translational periodicity, and we cannot define anything
like a "phonon band dispersion". Nevertheless, the concept of transmission channels is still
valid. As it was shown by Landauer \cite{Landauer}, no matter how scatterers are distributed in the system,
one can always calculate a transmission function, $\Xi(\omega)$, which describes the propagation
of quasiparticles between two reservoirs connected to the system, with different chemical
potentials. In contrast to translationally periodic systems, where the conduction channels
always yield integer quanta of transmission, now the waves can be partially reflected 
resulting in non integral transmissions. Thus in this case, instead of Eq.~(\ref{eq:sigmasat}), we have
\begin{eqnarray}
\sigma = \int_{0}^{\omega^{fin}} \Xi(\omega) {\partial\over\partial
 T}
\left [{1 \over {e^{\hbar \omega / k_B T} - 1}} \right ]  \hbar\omega {d\omega\over 2\pi},
\label{eq:conductance}
\end{eqnarray}
where the phonon spectrum extends between frequencies $0$ and $\omega^{fin}$.
Eq.~(\ref{eq:conductance}) is of general validity and involves no approximations. 
(This equation is rigorously derived in Section~\ref{sec:totalcurrent}.)
In the next subsection we explain how $\Xi$ is calculated exactly for independent phonons. 

Scattering causes the conductance to vary 
with the nanowire's length. For 3-D
systems in the diffusive regime, the conductance varies inversely proportional to the system's length, $L$.
Because of this one usually defines a "conductivity", as
\begin{eqnarray}
\kappa = {L \sigma (L)\over s},
\end{eqnarray}
where $s$ is the sample's cross section.
However, for nanowires, the length dependence of $\sigma$ cannot be just assumed to have a $1/L$ form,
but needs to be calculated. For short $L$, $\sigma$ saturates to $\sigma_{sat}$. For long
$L$ the behavior depends on the type of scattering and the wire properties. The transition between
the two limits also constitutes an important problem (studied in Section~\ref{sec:results}).
In order to attack these issues, the transmission function has to be calculated for the
atomically described, non-periodical, infinite system. The formalism allowing to do this 
is explained in the next subsection.

\subsection{\label{sec:calc}Calculation of the transmission function}

\subsubsection{\label{sec:dynmatr}Interatomic potentials and the dynamical matrix}

The system's motion is determined by its dynamical matrix, obtained from the
interatomic potentials of the system \cite{Jones}.
The nondiagonal elements of the dynamical matrix ${\bf K}$
are calculated as
\begin{eqnarray}
k_{ij}={\partial^2 E\over \partial u_i \partial u_j}
\end{eqnarray}
where $E$ is the energy, and $u_i$ is the displacement of the $i^{th}$ degree of freedom with respect to 
its equilibrium value.
The diagonal elements are $k_{ii}=\sum_{j\neq i}-k_{ij}$ \cite{Economou}.

The dynamical equation of the system is
\begin{eqnarray}
(\omega^2{\bf M} + {\bf K})\bar{u} = \bar 0
\end{eqnarray}
where $\bf M$ is a diagonal
matrix with elements corresponding to the masses of the constituent atoms, and
$\omega$ is the vibrational frequency. This problem is analogous to that posed by a
non-orthogonal Hamiltonian in the case of electrons, if we replace the
energy, Hamiltonian, and overlap, by the square frequency, dynamical matrix, and mass matrix, respectively \cite{Mingo01,Goldberg}.

The presence of matrix $M$ instead of the identity matrix introduces mathematical
difficulties, if Green functions are used. The issue has been elegantly dealt with
in the theory of non-orthogonal tight binding \cite{Goldberg}:
the "non-orthogonal" problem is equivalent to the
"orthogonal" one
\begin{eqnarray}
(\omega^2{\bf I} + {\bf M}^{-1/2}{\bf K}{\bf M}^{-1/2})\bar{u} = \bar 0
\end{eqnarray}
The elements of the "orthogonalized" dynamical matrix elements are explicitly defined, as
\begin{eqnarray}
\tilde k_{ij}=-{k_{ij}\over \sqrt{M_i M_j}},
\end{eqnarray}
and the dynamic equation is recast as
\begin{eqnarray}
(\omega^2{\bf I} - {\bf \tilde K})\bar{u} = \bar 0,
\label{eq:dynamic}
\end{eqnarray}
which now can be treated using Green function methods.

\subsubsection{\label{sec:current}Local heat current density}

We are faced with the problem of calculating the local heat flow.
The following paragraphs derive the necessary equations. The 
section following this one will express all the equations in terms
of Green functions, so that there is no need to calculate 
individual wave functions.

Let us consider an arbitrary wave (not necessarily an eigensolution) $u_i(t)$, propagating in the system.
The total energy can be expressed as a sum over each
degree of freedom, $E=\sum_i E_i$, with
\begin{eqnarray}
E_i = -{1\over 2}\sum_j u_i k_{ij} u_j + {M_i\over 2}\dot u_i^2
\end{eqnarray}
Using $M_i \ddot u_i = -k_{ij} u_j$, the local change of energy with time is
\begin{eqnarray}
{d E_i \over d t} = {1 \over 2} \sum_j (\dot u_i k_{ij} u_j - u_i k_{ij} \dot u_j) \equiv \sum_j J_{ij}
\end{eqnarray}
The local current between each pair of local degrees of freedom is thus
naturally defined, as
\begin{eqnarray}
J_{ij} = {1 \over 2} (\dot u_i k_{ij} u_j - u_i k_{ij} \dot u_j)
\end{eqnarray}

For a given phonon of frequency $\omega$
we rewrite $u$ and $\dot u$ in terms of the complex
wave $\phi(t)\equiv \psi e^{i\omega t}$:

\begin{eqnarray}
u_i (t)= Re[\phi_i (t)]/\sqrt M_i \equiv \phi_i^R (t)/\sqrt M_i\\
\dot{u_i}(t)=-\omega Im[\phi_i(t)]/\sqrt M_i \equiv -\omega \phi_i^I(t)/\sqrt M_i 
\end{eqnarray}
Hence, the current associated to that particular phonon between the $i$ and $j$ local
degrees of freedom is
\begin{eqnarray}
J_{ij} = {1\over2} \omega (\phi_i^R \tilde k_{ij} \phi^I_j - \phi_i^I \tilde k_{ij}\phi_j^R) \nonumber \\
= {\omega\over 2} Im(\phi_i^* \tilde k_{ij} \phi_j)
\label{eq:jota}
\end{eqnarray}

The $\phi$ are solutions of the eigenvalue problem, Eq.~(\ref{eq:dynamic}). The normalization
condition for the phonon amplitude follows from equating the wave's energy to $\hbar\omega$, as
\begin{eqnarray}
\sum_i |\phi_i|^2 = 2 \hbar/\omega
\label{eq:normalization}
\end{eqnarray}

\subsubsection{\label{sec:totalcurrent}The total current}

\begin{figure}
\includegraphics[width=8.5 cm]{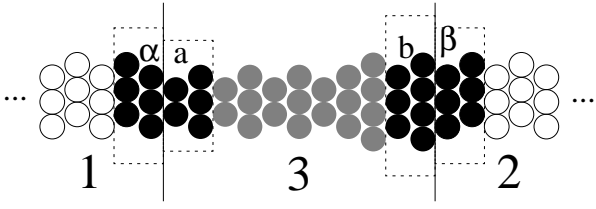}
  \caption{Nomenclature used in Section~\ref{sec:totalcurrent}.
The semi-infinite
leads 1 and 2 are projected onto subsets $\alpha$ and $\beta$, and
the total Green function for susbsystem 3 only needs to be computed between
atoms of subsets $a$ and $b$.
  }

\label{sketch}
\end{figure}

We now proceed to derive the expression for the total heat current given exclusively in terms of
Green functions. To this end, we subdivide the whole system into three regions, as indicated in
Fig.~\ref{sketch}. The two interfaces define four special groups of atoms labelled $\alpha$, $a$, $b$ and $\beta$, as
depicted in the figure.  An atom belongs to one of these groups if there is a non-zero element of 
the dynamical matrix linking it to another atom on the opposite side of the interface.
In what follows, a bar on top of a symbol, say $\overline\varphi$, indicates the one column matrix formed by the values
of $\varphi_i$ at each degree of freedom of the particular group of atoms considered. A parenthesis
denotes the subsystem: for example $\overline \varphi(a)$ is formed by values of $\varphi$ at the
atoms of subgroup $a$ only, and its dimension is $3$ times the number of atoms in $a$. Similarly,
a bar below the symbol, say $\underline\varphi$, denotes a one row matrix. We denote the traveling wave solutions
of the total system by $\overline\Phi_n$. The total current in terms of them is then given by (see Eq.~\ref{eq:jota})
\begin{eqnarray}
J = \sum_n N_n {\omega_n\over 2} Im[\underline\Phi_n^*(b){\bf \tilde k}_{b\beta}\overline\Phi_n(\beta)]
\end{eqnarray}
where $N_n$ is the number of phonons in state $n$.

Using standard scattering theory \cite{Economou} one can express the eigenfunctions of the total system,
$\Phi_n$ in terms of the retarded Green function matrix of the total system, ${\bf G}$, and the eigenfunctions of 
the decoupled system at the left of interface 1 (that is, with the coupling between $\alpha$ and $a$ set to zero).
Denoting these wavefunctions for the decoupled system as $\bar\phi(\alpha)$, one has for the waves propagating 
from $\alpha$ towards $\beta$,
\begin{eqnarray}
\bar\Phi_n(b)=&{\bf G}_{ba} {\bf \tilde k}_{a\alpha} \bar\phi_n(\alpha) \\
\bar\Phi_n(\beta)=&{\bf G}_{\beta a} {\bf \tilde k}_{a\alpha} \bar\phi_n(\alpha) = 
{\bf g}_{\beta \beta}{\bf \tilde k}_{\beta b}{\bf G}_{ba} {\bf \tilde k}_{a\alpha} \bar\phi_n(\alpha),&
\end{eqnarray}
where ${\bf g}_{\alpha\alpha}$ and ${\bf g}_{\beta\beta}$
are the retarded Green function matrices corresponding to the decoupled systems 
(i.e. with ${\bf \tilde k}_{\alpha a}={\bf 0}$ and ${\bf \tilde k}_{b \beta}={\bf 0}$).

Inserting these expressions in the equation for the current we have
\begin{eqnarray}
&&J=\nonumber\\
&&{Im\over 2} \sum_n \Delta N_{\omega}\omega_n \underline {\phi_n^*}(\alpha) {\bf\tilde k}_{\alpha a} {\bf G}_{ab}^* {\bf\tilde k}_{b\beta}
{\bf g}_{\beta\beta} {\bf \tilde k}_{\beta b} {\bf G}_{ba} {\bf\tilde k}_{a\alpha} \bar \phi_n(\alpha)\nonumber\\
&&={1\over 2}\sum_n \Delta N_{\omega}\omega_n \times \nonumber\\
&&\times Tr\left[
Im\{\bar\phi_n(\alpha) \underline {\phi_n^*}(\alpha)
 {\bf\tilde k}_{\alpha a}
{\bf G}_{ab}^*
{\bf\tilde k}_{b\beta}
{\bf g}_{\beta\beta} {\bf \tilde k}_{\beta b} {\bf G}_{ba} {\bf\tilde k}_{a\alpha}\}\right],
\nonumber\\
\label{eq:jaux}
\end{eqnarray}
where the sum extends to all states of the uncoupled subsystem 1, and 
\begin{eqnarray}
\Delta N \equiv N^+ -N^- = 
{\hbar\omega\over k_B T^2}{e^{\hbar\omega/k_BT}\over(e^{\hbar\omega/k_BT}-1)^2} \Delta T
\end{eqnarray}
is the occupation difference between phonons travelling forwards and backwards.

It is not feasible to explicitly compute the eigenfunctions $\bar\phi_n$ for a semi-infinite system.
This can be avoided by expressing everything in terms of Green functions, which allow to project 
semi-infinite systems on the atomic groups considered. 
From the normalization condition, Eq.~(\ref{eq:normalization}), and the well known relation
between the green function and the density of states \cite{Economou}, 
it follows that, at any infinitesimal frequency square interval,
$\{\epsilon,\epsilon+d\epsilon\}$,
\begin{equation}
\sum_{_{n (\omega_n^2\in\{\epsilon,\epsilon+d\epsilon\})}} \omega_n\bar\phi_n(\alpha)\underline{\phi_n^*}(\alpha)=
2{\hbar\over\pi}Im[{\bf g}_{\alpha\alpha}(\epsilon)]d\epsilon,
\end{equation} 
where $\epsilon\equiv \omega^2$.
Using this relation, the expression for the current, Eq.~(\ref{eq:jaux}), becomes
\begin{eqnarray}
J=&&\int_0^{\infty}{\hbar\omega\over 2\pi}\Delta N(\omega)\Xi(\omega) d\omega
\label{eq:current}\\
\Xi(\omega) =&& 4 Tr[{\bf\tilde k}_{a\alpha}Im[{\bf g}_{\alpha\alpha}] {\bf\tilde k}_{\alpha a} {\bf G}_{ab}^* {\bf\tilde
 k}_{b\beta}
Im[{\bf g}_{\beta\beta}] {\bf \tilde k}_{\beta b} {\bf G}_{ba}] \nonumber \\
\label{eq:transmission}
\end{eqnarray}
where the transmission function $\Xi(\omega)$ has been defined. The thermal conductance,
Eq.~(\ref{eq:conductance}), follows straightforwardly from Eq.~(\ref{eq:current}). Now we only need to compute the
Green functions ${\bf g}$ and ${\bf G}$, 
and then use Eqs.~(\ref{eq:transmission}) and (\ref{eq:current}) to calculate thermal currents.

\subsubsection{\label{sec:Green}Calculation of the Green functions for infinite systems without translational periodicity}

The system we are treating is infinitely extended in the heat propagation direction, but it does not have translational 
symmetry. Therefore, the calculation of the Green functions requires the use of projection techniques. The most 
efficient way to obtain the Green function of the semi-infinite systems 1 and 2 projected at subsystems $\alpha$ and 
$\beta$ is to use the decimation technique \cite{Guinea}, which is based in a renormalization procedure. After
the ${\bf g}_{\alpha\alpha}$ and ${\bf g}_{\beta\beta}$ are calculated in this way, one can calculate the total
retarded Green function of the system everywhere in subsystem 3 (Fig.~\ref{sketch}), as
\begin{eqnarray}
{\bf G}(\omega^2) = [\omega^2{\bf I} - {\bf \tilde K} - {\bf\Sigma}_1(\omega^2) -{\bf\Sigma}_2(\omega^2)]^{-1}
\label{eq:Green}
\end{eqnarray}
where the self-energy matrix ${\bf \Sigma}_1$ is defined as ${\bf \tilde k}_{\alpha a}{\bf g}_{aa}
{\bf \tilde k}_{a\alpha}$ on the elements belonging to subset $a$, and zero otherwise, and ${\bf \Sigma}_2$ is
defined similarly on subset $b$.

\subsection{\label{sec:SCflux}Self-consistent flux model for efficient computation of boundary scattering}

If we know the dynamical matrix,
the formalism explained in the previous paragraphs is capable of
accounting for any type of scattering mechanism other than anharmonicity (for the role played
by anharmonicity, see appendix).
It is straighforward to include isotopes by varying the mass of given atoms; substitutional
impurities may be introduced by varying both the mass of selected atoms and the
links to their neighbors; vacancies would be included by removing single atoms and allowing 
the system to relax; etc. We might thus be tempted to treat boundary scattering 
in this same spirit, by including surface adsorbates or imperfections that can lead
to effective boundary scattering in the nanowire. Adding single adsorbed impurities is a 
rather simple matter, but, could one consider an amorphous overlayer on the
same footing? For example, it is known that Si nanowires are covered by an amorphous $SiO_2$
overlayer, several $nm$ in thickness \cite{Lieber}, and it is important to know how this overlayer affects 
the phonon scattering. A brute force approach would be to input the atomic positions of all 
constituents in the silica overlayer, as well as the wire, and then calculate the dynamical
matrix of the whole system and compute its transmission.

\begin{figure} 
\mbox{\psfig{figure=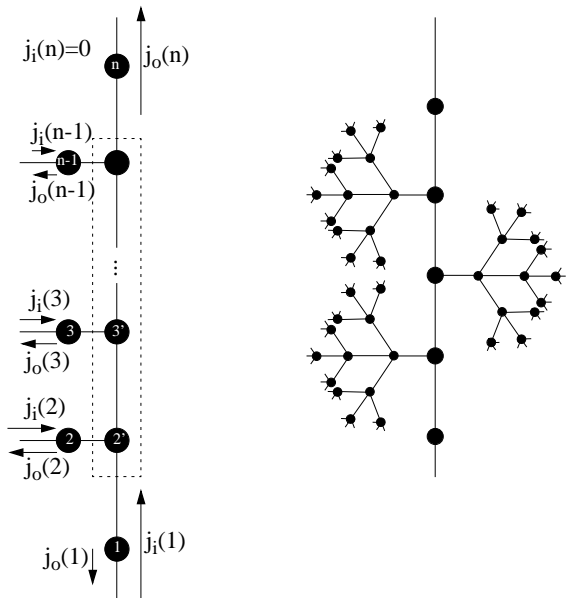,width=8.25cm}}
\caption{Left: scheme of a linear chain system showing the
nomenclature used in the self consistent flux technique of Subsection~\ref{sec:SCflux}.
Right: scheme of 1-D wire with Bethe lattices of coordination
4 attached.
}

\label{chain} 
\end{figure}

This procedure, however, is a nearly impossible task in practice. The added overlayer
considerably increases the size of the system, and it also results in an extremely irregular transmission function,
which has to be computed
at a very large number of different frequencies.
The integration of such a transmission in Eq.~(\ref{eq:conductance}), consequently, becomes cumbersome.
To solve these difficulties,
we have developed the self-consistent flux model. This model's basic idea consists in taking the
limit of an infinitely thick overlayer. In the same spirit of the Bethe lattice approach,
as used for amorphous materials \cite{Economou}, we consider the nanowire connected to Bethe lattice branches of the
amorphous material (see Fig.~\ref{chain}). The  phonon mean free path in amorphous materials is of the 
order of the interatomic spacing \cite{Klemens}. Therefore, the lack of connectivity in the Bethe lattice
approach does not represent a problem. The adequacy of the Bethe lattice approach for the study of 
amorphous systems has been investigated in refs.~\onlinecite{Verges,Laughlin}.

When the overlayer is substituted by an infinitely thick one, an essential difference appears.
Now, the self-energies associated to the overlayer do not correspond to clusters. Therefore,
they no longer are purely real, but contain a finite continuous imaginary part. In other words,
if one calculates the transmission of the system by Eq.~(\ref{eq:transmission}), part of the phonon flux will be 
effectively lost through the side branches, since now they are infinite and allow for the 
current to go away without returning. However, this is not physically correct, for we know that
the overlayer is finite and the heat current must eventually come back to the wire and escape 
through the contacts. In a steady state, this means that we must allow for an equivalent
ammount of phonon flux to re-enter the system through each of the side branches.
One may find a self-consistent solution to this problem
in the form of an iterative procedure. This is indeed not necessary,
for the self-consistent solution can be directly found by an algebraic method, as we show below.

To clarify the nomenclature in the following discussion,
a general scheme of the nanowire with part of its length coated is shown in Fig.~\ref{chain}, left.
The represented nanowire is three dimensional, thus each black dot in the figure stands for
a group of atoms, and each line joining dots stands for a set of interatomic links.
The dotted line encloses the wire's groups of atoms that are coated. Bethe lattices are attached
to those groups of atoms, as shown in the figure. Phonons enter the system through group $1$, and
exit at the other lead through group $n$.
First of all, we obtain the partial transmissivities, $\cal T$, between every pair of branches in the
system, using Eq.~(\ref{eq:transmission}),
\begin{eqnarray}
{\cal T}_{ij} = 
{4\over N} Tr[{\bf\tilde k}_{ii'}Im[{\bf g}_{i'i'}] {\bf\tilde k}_{i'i} {\bf G}_{ij}^* {\bf\tilde
 k}_{jj'}
Im[{\bf g}_{j'j'}] {\bf \tilde k}_{j'j} {\bf G}_{ji}]
\label{eq:partialcond}
\end{eqnarray}
where $i$ and $i'$ label neighboring atomic subsets belonging to the branch and wire respectively,
for the first branch of the pair, and equivalently $j$ and $j'$ for the second branch.
$N$ is the number of degrees of freedom in node $i'$ or $j'$. (All nodes are assumed to
contain the same number of atoms for simplicity.)

We recall that each link to a branch in Fig.~\ref{chain} may comprise interactions with
several atoms, in the three geometrical directions, so the Green functions above are 
matrices. The transmissivity from one branch to itself is not defined by the 
above formula. Instead, we must define it as the reflectivity,
\begin{eqnarray}
{\cal T}_{ii} = 1 - \sum_{j \neq  i}{\cal T}_{ij}.
\label{refl}
\end{eqnarray}

We define the "injected" and "outflowing" flux arrays, $\bar{j}_{i}$ and $\bar{j}_{o}$ as the
fluxes coming "into" and "out of" the system via its surrounding nodes (see Fig.~\ref{chain}). The total flux array, $\bar j$,
is then 
\begin{eqnarray}
\bar{j} = \bar{j}_i - \bar{j}_o
\end{eqnarray}
and it must fulfill
\begin{eqnarray}
j(1) = j_i(1) - j_o(1) = j_o(n)
\label{eq:one}
\end{eqnarray}
because all the flux coming in through node $1$ must finally exit through node $n$.
Also, there is no flux injection from node $n$, so
\begin{eqnarray}
j_i(n)=0
\label{eq:three}
\end{eqnarray}

All the other modes must fulfill
\begin{eqnarray}
j(l) = j_i(l) - j_o(l) = 0, \forall l \neq 1,n
\label{eq:four}
\end{eqnarray}

We can express Eqs.~(\ref{eq:one},\ref{eq:three},\ref{eq:four}) in matrix form as
\begin{eqnarray}
j_i = {\bf A} j_o
\end{eqnarray}
with
\begin{eqnarray}
{\bf A}=\pmatrix{
1 & 0 & 0 & \cdots & 0 & 1 \cr
0 & 1 & 0 & \cdots & 0 & 0 \cr
0 & 0 & 1 & \cdots & 0 & 0 \cr
\vdots&\vdots &\vdots& \ddots &\vdots& \vdots   \cr
0&0&0&\cdots &  1 & 0 \cr
0&0&0&\cdots &  0 & 0 \cr}
\end{eqnarray}

On the other hand, the outflowing currents are related to the injected ones by
\begin{eqnarray}
\bar j_o = {\bf\cal T} \bar j_i
\end{eqnarray}
which implies
\begin{eqnarray}
\bar j_o = {\bf\cal T} {\bf A} \bar j_o
\end{eqnarray}

Therefore, the self-consistent array $\bar j_o$ is proportional to the
eigenvector $\bar v_1$ of ${\bf {\cal T}A}$ with eigenvalue $1$, such that 
${\bf {\cal T}A} \bar v_1 = \bar v_1$. 
The existence of eigenvalue $1$ is guaranteed by the
way we have defined ${\cal T}_{ii}$ (Eq.~(\ref{refl})) and by the fact that columns $1$ and 
$n$ in matrix ${\bf {\cal T} A}$ coincide.

The transmission corresponds to the outgoing flux at node $n$
when there is one unit of flux inciding per degree of freedom in node $1$.
Thus, imposing $j_i(1) \equiv N = j_o(1) + j_o(n)$, where the second equality
follows from Eq.~(\ref{eq:one}), gives
\begin{eqnarray}
\bar j_o = {N \over v_1(1)+v_1(n)}\bar v_1
\end{eqnarray}
and the self-consistent transmission is

\begin{eqnarray}
\Xi = j_o(n) = N{v_1(n) \over v_1(1)+v_1(n)}
\label{eq:SCtransmission}
\end{eqnarray}

It can be easily verified that for the uncoated case the calculation using Eq.~(\ref{eq:SCtransmission}) coincides
with that of Eq.~(\ref{eq:transmission}).

In the next section we explicitly demonstrate that the SCFlux approach yields 
the same result that the calculation using large clusters attached to the
wire. Therefore, the SCFlux technique is of tremendous importance, since it enables
to study the otherwise untractable problem of thermal transport through
thick-amorphous-coated real wires.

\section{\label{sec:results}Results}

New physics regarding how the amorphous coating influences the thermal
transport through a wire can now be learnt using the above method:
Does the coating affect all frequencies equally, or are there important differences in the
scattering as a function of frequency? Does diffusive transport arise? And, if so, how long a segment
has to be coated in order for transport to diffusive rather than ballistic? How does transport depend on
the coupling at the wire-coating interface? Etc. 

In the next two subsections we present results for two different systems. First, a one dimensional model, which
is simple enough to allow for a thorough investigation without encountering size limitations, and 
at the same time displays many general features that apply to real systems.
Afterwards, we present results for real 3-D silicon nanowires.

\subsection{\label{sec:1dwire}Study of a one dimensional model}

A wide range of interesting phenomena is obtained already in the case
where there is only one degree of freedom per atom,
in a linear chain. Bethe lattices with also one degree of freedom per
atom are attached to a section of the chain. 
The wire's properties are determined by the value of a spring constant
$K_W$. For the Bethe lattices, we assume them to be linear chains (i.e. $Z=2$ in
the nomenclature of ref. \onlinecite{Economou}), and they are characterized  by their spring constant
$K_B$. The third constant involved is the coating-wire interaction, $K_{BW}$,
connecting the end of the Bethe lattice to the wire.

\subsubsection{\label{sec:compare}Comparison between SCFlux and cluster calculations}

\begin{figure}
\mbox{\psfig{figure=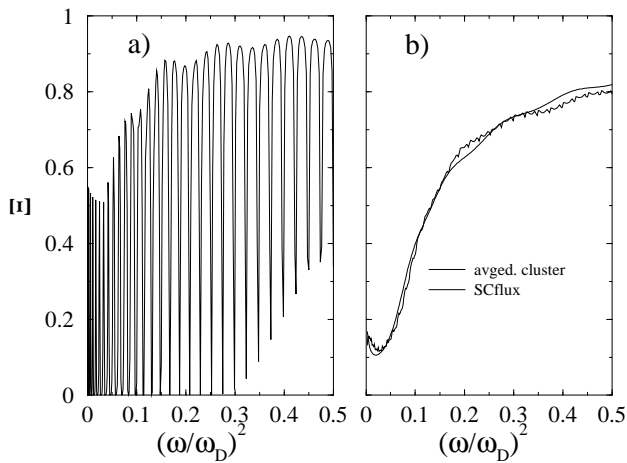,width=8.25cm}}
\caption{Comparison between the transmission calculated
by using finite clusters as scatterers or the SCFlux
method with infinite Bethe lattices. a) Cluster transmission. (10 clusters
comprising 60 atoms each are attached to the chain.) b)
Thick line: SCFlux transmission;
thin line: the curve in "a)" averaged in a range of
frequencies around each point.
As more atoms are added to the
clusters, the averaged curve approaches the SCFlux result
more closely.
}
\label{cluster}
\end{figure}

First of all, we provide a particular example that demonstrates the equivalence between the SCFlux calculation, and 
the much more computationally demanding calculation using large finite clusters
for the coating. The transmission was computed for a system with ten clusters
attached to it at ten consecutive atoms. Each of the clusters is composed of 
sixty overlayer atoms. The transmission, calculated via Eq.~(\ref{eq:transmission}), is shown in Fig.~\ref{cluster}(a).
We shall abbreviately refer to this curve as the "cluster transmission".
A large number of closely spaced antiresonances occurs, due to the finite size of
the clusters. Thus, the curve had to be computed at many different frequency points. 
\footnote{As it is customary in Green function cluster calculations,
a small imaginary part, $i\delta$, is added to the frequency to smoothen singularities. To avoid
spurious effects, the imaginary part has to be smaller when the density of singularities increases
near the band edges, so $\delta\propto \sqrt{2-\cos^2\tilde\omega}/\sqrt 2$ was used.}

The transmission given by the SCFlux method is shown as the
thicker line in Fig.~\ref{cluster}(b). We shall refer to it as the "SCFlux transmission".
Its smooth shape has little visual resemblance with the
cluster transmission.

We can now compute
the thermal conductance as a function of temperature,
using both the cluster and SCFlux transmissions, and compare the results.
We define the dimensionless
thermal conductance as
\begin{eqnarray}
\tilde\sigma = {6\over\pi} {\sigma\over k_b\omega_D}
\label{eq:nondimenconductance}
\end{eqnarray}
where $\omega_D=2K_B$ is the Debye frequency of the infinite 1-D atomic chain.
Using Eq.~(\ref{eq:conductance}), the dimensionless conductance can be evaluated in terms of the dimensionless
temperature $\tilde T\equiv {k_b\over\hbar\omega_D}T$, and the dimensionless
frequency $\tilde\omega\equiv{\omega\over\omega_D}$, as
\footnote {For $\tilde T\rightarrow 0$ the dimensionless conductance tends to $\tilde\sigma \rightarrow 
\Xi(0)\tilde T$. This means that for $\Xi(0)=1$ the conductance becomes 
${k_b^2\pi^2\over 3h}T$, i.e. one quantum of thermal conductance. Also, for a constant
transmission $\Xi(\tilde\omega)=\Xi_0$, we have $\lim_{\tilde T\to \infty}\tilde\sigma
={3\pi^2}\Xi_0$.}
\begin{eqnarray}
\tilde\sigma={3\over\pi^2}\int_0^1 \Xi (\tilde\omega){\tilde\omega^2\over\tilde T^2}
{e^{\tilde\omega/\tilde T}\over(e^{\tilde\omega/\tilde T}-1)^2} d\tilde\omega
\end{eqnarray}

\begin{figure}
\mbox{\psfig{figure=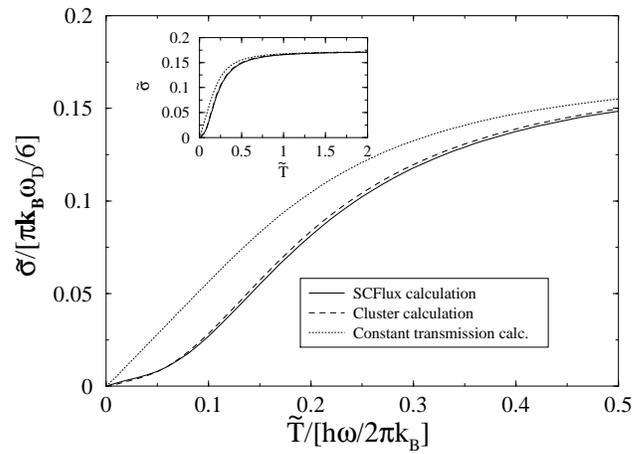,width=8.25cm}}
\caption{Comparison between the dimensionless thermal conductance as a function of the dimensionless
temperature, calculated from the {\it cluster} transmission (dashed line),  and from the {\it SCFlux}
transmission (solid line). Although the two transmissions have very different shapes (c.f.Fig.~\ref{cluster}), the
conductances obtained from them are nearly the same. The conductance obtained from a {\it constant} transmission
function is also shown (dotted line) to stress the importat role played by the frequency
dependence of the transmission function.
}
\label{conductance}
\end{figure}

The dimensionless conductance calculated from the cluster transmission is shown as the
thick dashed line in Fig.~\ref{conductance}, as a function of the dimensionless temperature.
Now this curve can be compared with the thermal conductance calculated using
the SCFlux transmission. 
Remarkably, the latter yields the thick solid curve in Fig.~\ref{conductance}: almost exactly the same thermal 
conductance curve than that computed from the cluster
transmission. The reason why the two apparently very different transmissions
yield virtually the same conductance can be better understood by averaging the cluster transmission in 
a range of frequencies, so as to smooth out its structure. We have convolved the curve in Fig.~\ref{cluster}(a) with a 
truncated Gaussian kernel extended $\pm\omega_D/60$ around each point. The cluster transmission
averaged in this way is shown as the thin line in Fig.~\ref{cluster}(b), and is very close to the SCFlux transmission.

Thus we see that the transmission calculated with the SCFlux method does indeed yield the same thermal conductance 
as if we include the thick coating in the form of large finite clusters. But the SCFlux method requires 
a much smaller computation, because it yields a smooth transmission curve, while the cluster calculation
needs to evaluate a far larger number of points in order to account for the closely spaced antiresonant
structure associated to the finite clusters.

To quantify
the difference between the {\it real} conductance and the one obtained using a {\it frequency independent} boundary scattering,
the latter is also plotted in Fig.~\ref{conductance}, as the thin dotted curve. The value of the constant transmission has
been fixed to yield the same limiting conductance in the $\omega\to\infty$ limit
as the real conductance curves (see inset of Fig.~\ref{conductance}). As the figure shows, important differences between the
oversimplified constant transmission model and the real transmission calculation exist
up to well above the debye temperature.

\subsubsection{\label{sec:trfunc}Study of the transmission function}

Now we proceed to study the behavior of the transmission as a function of 
the different physical parameters in play. More concretely, we want to know
how the transmission depends on frequency $\omega$, coated length $L$, 
and strength of the interface spring constant $K_{BW}$.
The analysis will show us how the system goes from ballistic to diffusive 
behavior, as a function of these physical parameters.

We set $K_W$ to be the unit of spring strength. The mass of all atoms,
$M$, is defined as $1$ also. Frequencies are in units of $\sqrt{K_W/M}$.
The coated length $L$ is given in lattice units, being equivalent to the
number of Bethe lattices attached. 

\begin{figure}
\mbox{\psfig{figure=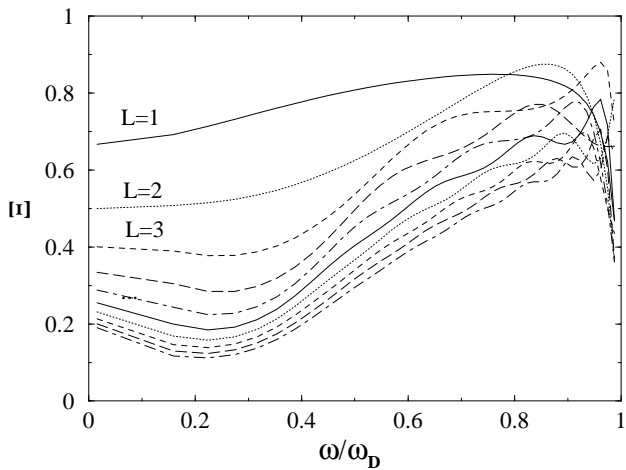,width=8.25cm}}
\caption{SCFlux transmission as a function of frequency
for infinite 1-D chain, for coated segments of different
lengths, L (= number of lattices attached).
}

\label{trvsfreq}
\end{figure}

First let us take a look at the behavior of the transmission function with
frequency, in Fig.~\ref{trvsfreq} (for $K_W /K_B = 1$ and $K_{BW}=1$).
We notice that the effect of boundary scattering is to reduce the 
transmission rather uniformly throughout the whole frequency range. However,
it is apparent that the usual assumption of a constant scattering rate 
\cite{Klemens,Callaway,Holland,Asen-Palmer} is only
a rough approximation to the real scattering in nanowires. The low frequency limit behavior
displays an important feature: 
for wires where the coated segment is long enough ($n>3$ in the case shown),
the transmission near $\omega \simeq 0$ decreases with increasing frequency. This appears to be the
general case, as we will see in the next section. It has a direct physical
consequence: the slope of thermal conductance at very low temperatures
decreases below its $T=0$ value when temperature increases. This fact
has been shown to happen also in the case of surface roughness at non-coated wires \cite{Santamore}
(the well known "dip" of ref.\onlinecite{Schwab}.) We explicitly show this for Si nanowires in the 
next subsection.

In Fig.~\ref{trvsfreq} we have seen that the shape of the transmission function varies as a function of the 
coated length. Does it attain a limiting shape? And, how is this limit reached? This is the
subject of the {\it diffusive-ballistic transition}. The way it takes place will become clear from the
study of the transmission as a function of coated length, $L$. In principle, it is not obvious what
type of length dependence one could expect. For pure disorder, for example, an exponential
decrease in transmission is obtained in 1-D systems due to localization \cite{Beenaker}. As another example,
for wires with rough edges, an anomalous
$L^{-1/2}$ thermal conductance has also been reported under certain conditions \cite{Kambili}.
For amorphous-coated wires, the asymptotic behavior turns out to be $\sim L^{-1}$, as we now show.

\begin{figure}
\mbox{\psfig{figure=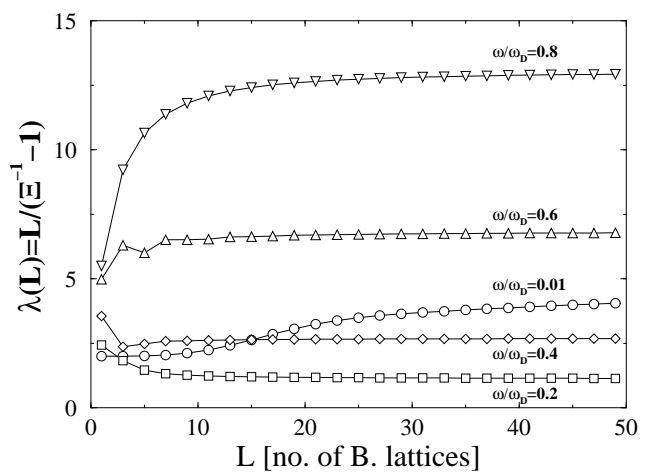,width=8.25cm}}
\caption{Plot of $\lambda(L)\equiv (\Xi_0/\Xi-1)^{-1}L$ as a function of the coated
length $L$, for different frequencies.
$\lambda(L)$ asymptotically converges to a limiting relaxation length value, $\lambda_{\infty}(\omega)$ for
long enough $L$.
}
\label{lambda}
\end{figure}

Analogously to ref.~\onlinecite{deJong}, we write the transmission as
\begin{eqnarray}
\Xi=\Xi_0(1+L/\lambda(L))^{-1}
\end{eqnarray}
This is merely a definition of $\lambda(L)$,
and it does not impose any limitations on the form of the transmission.
For the present 1-D system, the uncoated wire's transmission, $\Xi_0$, is equal to $1$ at all frequencies.
Now, the form of $\lambda(L)$, plotted in Fig.~\ref{lambda},
tells us the length dependence of the transmission. New physical insight can be
gained from this figure. First, we see that $\lambda(L)$ quickly tends to an asymptotic value. This
means that, for amorphously coated wires,
the long $L$ behavior of the transmission follows an $L^{-1}$ dependence \footnote{This can be understood
as the presence of the amorphous coating giving the wire a three dimensional character and preventing
localization.}.
More interestingly, the length it takes $\lambda(L)$ to saturate to its asymptotic value
becomes longer as the frequency gets closer to zero. Physically, this implies that the 
lowest frequency phonons can still travel nearly ballistically for system lengths at which higher
frequency phonons are transported in an almost totally diffusive fashion. 
In a practical measurement of nanowire thermal conductivity, this might lead to a wrong estimation
of the low temperature thermal conductivity, if the wire were not long enough.

\begin{figure}
\mbox{\psfig{figure=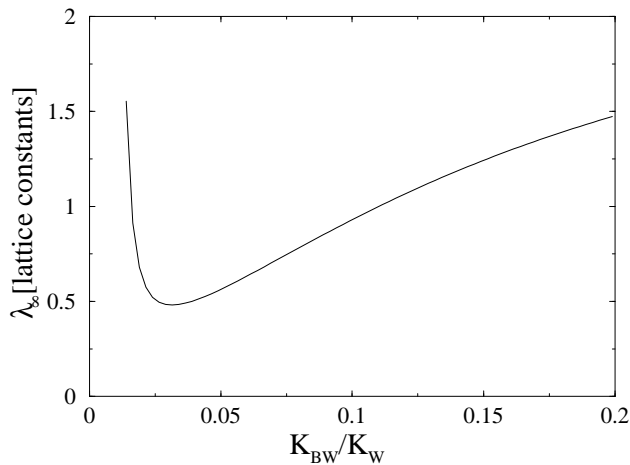,width=8.25cm}}
\caption{Phonon relaxation length $\lambda_{\infty}$ as a function of the
coupling between the wire and its coating lattices.
It reaches a minimum and then increases, asymptotically
approaching a limiting value. (In the particular case shown here, $\omega=0.05\omega_D$, and $\lambda\to 2.7$.)
}
\label{condvsstrength}
\end{figure}

In the diffusive regime, we can define a frequency dependent relaxation length, $\lambda_{\infty}$, as the limit
\begin{eqnarray}
\lambda_{\infty}(\omega)\equiv lim_{L\to\infty}\lambda(\omega). 
\label{eq:meanfp}
\end{eqnarray}
A very interesting phenomenon arises when we
study how this relaxation length depends on the strength of the spring constants coupling 
the wire to the amorphous coating.
We numerically computed $\lambda_{\infty}$ at frequency $\omega=0.05\omega_D$,
as a function of the wire-coating coupling strength. The result is plotted in Fig.\ref{condvsstrength}.
For decreasing values of the wire-coating coupling, $K_{BW}$,
the relaxation length quickly increases,
as expected. However, if the coupling is made stronger,
$\lambda_{\infty}$ reaches a minimum 
and then it
increases again as the coupling is increased, approaching
an asymptotic value. 
The existence of this {\it minimum relaxation length} can be interpretted
in microscopic terms: there is an optimum value of the coupling
that maximizes the heat-flow exchange to and from the coating, thus maximizing the scattering. For
very weak couplings the heat-flow exchange reduction is intuitively
obvious. For very strong couplings on the other hand, 
the linked atoms would behave more like
a rigid cluster, thus acting as a hard wall that also confines the phonons to
the wire rather than let them enter the amorphous region.
From a macroscopic point of view, the relaxation length
is minimum when the specularity factor is 0, in the
Casimir limit \cite{Ziman}. The minimum of $\lambda_{\infty}$ obtained
thus translates to this limit, and specularity 
increases in both ways around this point as a function
of $K_{BW}$.

\subsection{\label{sec:sinanow}Results for Si nanowires}

The advantage of the atomistic description used in this paper is that we can study wires made of "real" materials.
In other words, we are dealing with the real phonon spectrum of the system: the bands are dispersive, and their
frequency extension is finite, in contrast with the infinitely extended spectrum in a continuum elastic system.

\begin{figure}
\mbox{\psfig{figure=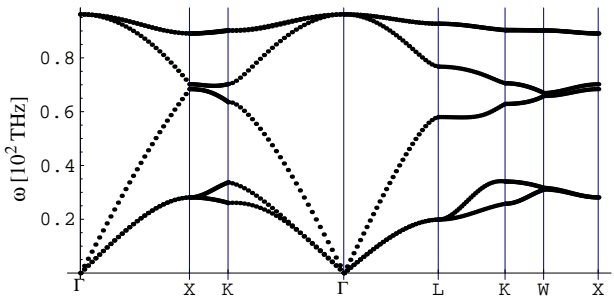,width=8.25cm}}
\caption{Phonon dispersion relations calculated for
bulk Si using Harrison's potential.
}
\label{disprel}
\end{figure}

We proceed to study the transmission of phonons in real Si nanowires, with 
an amorphous material coating the wire. 
The phonon dispersion relations of Si are fairly well reproduced
with an interatomic potential that includes only two and three body terms.
We use Harrison's potential \cite{Harrison}. 
The adequacy of this potential for Si is assesed by
the satisfactory dispersion relations calculated with it.
We show them for bulk Si in Fig.~\ref{disprel}. Despite having only two
parameters,
the shape and position of the bands are in reasonably good agreement with 
experimental results \cite{Siband,Weber}.

The two body contribution to the energy is
\begin{eqnarray}
\delta E_0 (i,j)={1\over 2}C_0 {(d_{i,j} - d_0)^2 \over d_0^2}
\end{eqnarray}
for every pair of nearest neighbors $i$ and $j$, where $d_{i,j}$ is the distance
 between the atoms and
$d_0$ is the lattice equilibrium distance. The three body contribution is
\begin{eqnarray}
\delta E_1(i,j,k)={1\over 2}C_1 \delta \Theta_{i,j,k}^2
\end{eqnarray}
for every pair of bonds joining atoms $i$, $j$ and $k$,
where $\delta\Theta$ is the deviation with respect to the
equilibrium angle between the two bonds in the lattice. Constants $C_0=49.1eV$ and
$C_1=1.07eV$ are taken from table 9-1 of ref.~\onlinecite{Harrison}.

\begin{figure}
\mbox{\psfig{figure=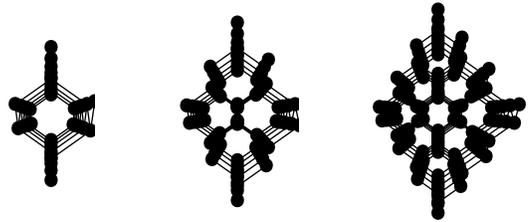,width=8.10cm}}
\caption{Cross sections of the three Si nanowires
studied: "2x2", "3x3" and "4x4".
}
\label{crosssect}
\end{figure}

\begin{figure}
\mbox{\psfig{figure=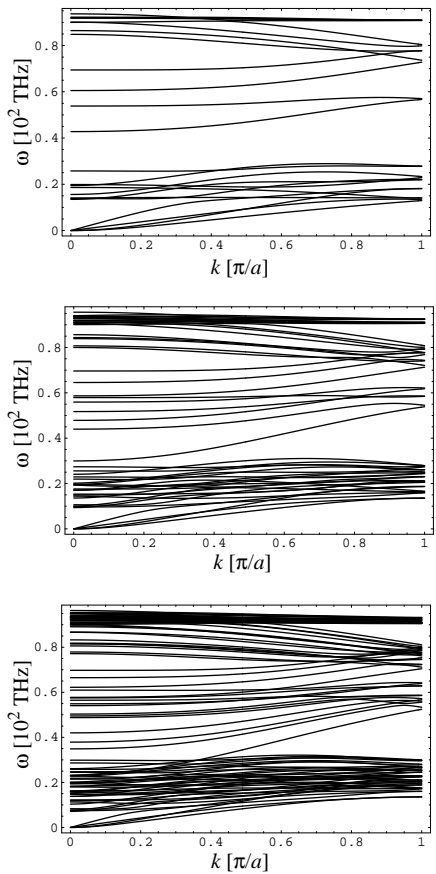,width=6.25cm}}
\caption{Phonon dispersion relations for the three Si
nanowires in Fig.~\ref{crosssect}.
}
\label{wdisp}
\end{figure}

For the coating we again consider Bethe lattices connected to each vibrational 
coordinate of the boundary atoms. The parameters of the Bethe lattice are chosen
to reproduce the frequency range of the known d.o.s. of silica \cite{Silica}, i.e. $2\times 10^{14}$ Hz.
A coordination number $Z=4$ is assumed.
We have considered (011) wires of three different widths, 
shown in Fig.~\ref{crosssect}.
The wires are coated by attaching Bethe lattices only to the outermost atoms. 
For the 2x2 wire all atoms are at the surface. The 3x3 wire has 16 surface atoms and 2 core atoms.
The 4x4 wire has 24 surface atoms and 8 core atoms.
The phonon dispersion relations for each of them is shown in Fig.~\ref{wdisp}. 

The procedure for the calculation is as follows. First, the nanowire lattice is
defined, and the dynamical matrix is obtained for the periodic unit cell, and 
for the link between two neighboring unit cells. The Bethe lattices are
included as a self-energy projected onto their last atom. The self-energy
due to the uncoated parts of the wire is calculated in the way of ref.~\onlinecite{Guinea}.
Then the Green function of the system is calculated (Eq.~(\ref{eq:Green})). The partial transmissivity
between each pair of branches (Eq.~(\ref{eq:partialcond})) is then calculated, 
and the SCFlux transmission (Eq.~(\ref{eq:SCtransmission})) is computed.

\begin{figure*}
\mbox{\psfig{figure=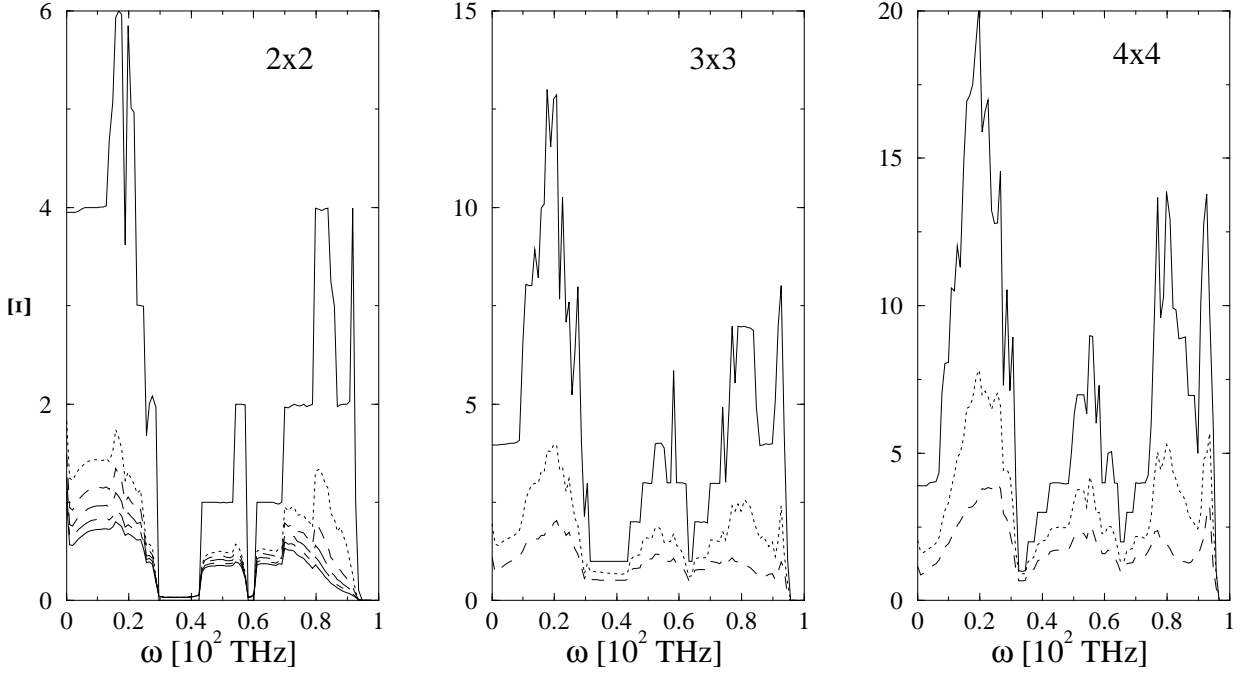,width=16.5cm}}
\caption{SCFlux transmission for Si nanowires. Left: transmission of the 2x2 wire for
0, 1, 2, 3, 4, and 5 unit cells coated. Center: transmission for the 3x3 wire with 0, 1, and 5
unit cells coated. Right: same, for the 4x4 wire.
}
\label{sitrans}
\end{figure*}

We show how the phonon transmission depends on frequency, for different values of the coated
length. Fig.~\ref{sitrans} shows $\Xi(\omega)$
as the coated length increases from
0 to 5 coated unit cells. 
In these curves,
all three spatial directions of each of the surface atoms are attached to Bethe lattices. 
The qualitative behavior is an overall uniform decrease of the
transmission curve, that maintains the peaks and curve features of the uncoated wire. This
is in agreement with the usual assumption of a constant scattering rate. However, quantitatively
the boundary scattering rate is not frequency independent, as it is apparent from the relative
heights of the peaks at different coating lengths.
In the absence of coating, the transmission curve at the lower end of the frequency spectrum
has a value of $4$. This is always the case, since there are four lowest frequency branches at
all nanowires: one dilatational, one torsional and two flexural branches \cite{branches}. 
The two flexural branches have a quadratic rather than linear
dispersion, as we can see in Fig.~(\ref{wdisp}), being a
consequence of the soft character of those modes. 

The transmission at the lowest frequency decreases more slowly than
that around the first peak ($\sim 20 THz$), as the coated length becomes larger. Thus scattering
is weaker for the lowest lying modes than for the bulk of transverse modes. This results in 
a flattening of the transmission as $L$ increases. The narrower the wire, the more pronounced
this effect becomes.

\begin{figure}
\mbox{\psfig{figure=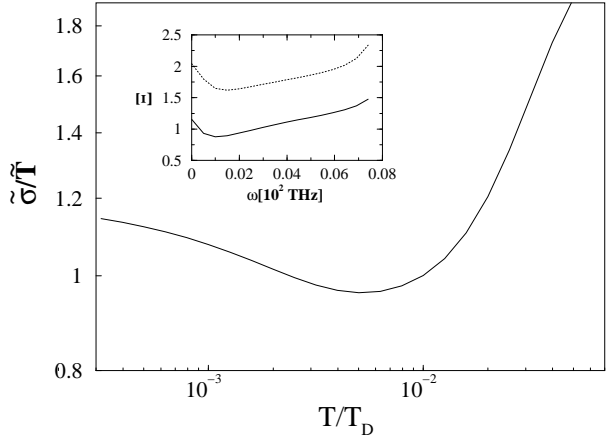,width=8.25cm}}
\caption{Low temperature behavior of the thermal conductance divided by temperature, for the 4x4 nanowire
with 5 unit cells coated, showing a dip similar to the one studied in \onlinecite{Santamore}. 
(The ordinate axis unit is ${6\over\pi}\hbar/k_B^2$.)
Inset: detail of the transmission function $\Xi(\omega)$ at low frequencies, for
the 4x4
wire with 1 (dotted line) and 5 (solid line) unit cells coated.
}

\label{lowtemp}
\end{figure}

The low frequency behavior of the transmission function also has important physical implications.
Similarly as what we saw in the 1-D example of Section~\ref{sec:1dwire}., for real Si nanowires the transmission
function also decreases below its $\omega=0$ value for small values of the frequency,
and increases again as the frequency rises further (see inset in Fig.~\ref{lowtemp}). 
The result is that, at low, finite temperatures, the thermal conductance divided by $T$
decreases with respect to its zero temperature value, thus 
displaying a dip (see Fig.~\ref{lowtemp}). A dip in $\sigma(T)/T$ has been reported in ref.~\onlinecite{Schwab}, and theoretically
explained in ref.~\onlinecite{Santamore} to be the result of surface roughness scattering. In our case of amorphous-coated
nanowires we see that an analogous dip occurs, although the system considered here is basically different from
that one, and the conductance reduction phenomenon cannot be considered to be the same one. Some essential
differences exist in the case of scattering by amorphous coating with repect to the roughness scattering case.
One is that, at zero frequency, the scattering rate is zero in the case of roughness scattering,
while it is different than zero in the case of amorphous-coating scattering. As a result, it is possible
to measure a conductance close to 4 times the quantum of conductance at uncoated wires, but for 
wires with a thick amorphous coating the measurement will in general yield smaller values at the very low temperature limit.
Another difference pointed out earlier is the exponential decrease of conductance with length in the 
rough wires, as compared with the inversely proportional dependence in the long length limit for amorphous-coated wires.

\begin{figure}
\mbox{\psfig{figure=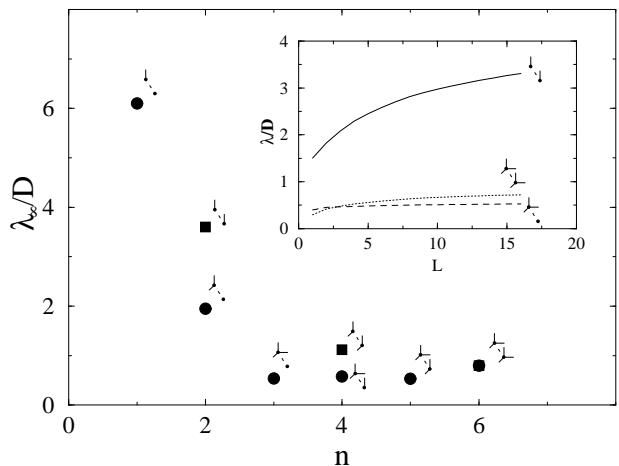,width=8.25cm}}
\caption{Phonon relaxation length $\lambda_{\infty}$ in 4x4 wire
at frequency $\omega=16$ THz,
for different ways of attachment between the coating and the wire
surface atoms, in units of the wire's diameter $D$. The scheme near each datapoint represents
the directions in which the Bethe branches are attached to each surface pair of atoms. The inset shows $\lambda$ versus the
coated length $L$ (in unit cell lengths), for three particular attachment configurations.
}
\label{meanfreepath}
\end{figure}

We have also calculated the relaxation length, $\lambda_{\infty}$ (Eq.~(\ref{eq:meanfp})),
as a function of the Bethe lattice coverage, in Fig.~\ref{meanfreepath}.
The abscissa corresponds to the number of Bethe lattices attached to each pair of surface
atoms of the 4x4 nanowire. Since there are three spatial directions, we can attach 
up to 6 lattices to each of the pairs. We show $\lambda_{\infty}$ as a function of
the number of directions attached, for different ways of attachment. The directions in which the Bethe lattices are 
attached, and the number of them, are depicted schematically near the points in the graph.
The calculation shows that the relaxation length is strongly dependent on the character of the
bonding to the coating material. For a densely covered wire, the relaxation length becomes 
of the order of the wire's core diameter. On the other hand, if the surface bonds to the 
coating are less dense, with one dominant direction and two much weaker links in the other
two directions, then the relaxation length can attain values between 2 and 4 times the diameter.
When there is only partial bonding of part of the surface atoms, the relaxation length increases
further. 
The way in which $\lambda$ approaches its saturation value, $\lambda_{\infty}$ is shown in the inset of Fig.~\ref{meanfreepath}.
With only one link per surface atom (solid curve), it is necessary to coat a considerable length in order to 
attain the diffusive limit. With more interface links, the diffusive limit is reached at shorter coated lengths.
It is observed that, not only the number of links, but also their orientation, has an influence in the 
relaxation length. 

A larger number of interface links does not necessarily imply a shorter relaxation length.
A fully covered surface with three links per atom, for example, displays a larger
relaxation length than a surface with only half of the surface atoms directly attached to the coating (Fig.~\ref{meanfreepath}). 
This is another manifestation of the phenomenon noted in Subsection~\ref{sec:1dwire}, where we saw that a very
strong interaction between wire and coating can result in longer relaxation lengths.

\section{\label{sec:conclusions}Conclusions}

We have presented a formalism to calculate the phonon transmission function of 
an atomically described system. With this formalism we have studied the
problem of phonon transport in nanowires with part of their length coated by an amorphous material.
A self-consistent approach has been developed that
uses Bethe lattices to treat the amorphous material. The efficiency of the approach,
and its equivalence to a much more computationally demanding cluster calculation, have been
explicitly illustrated via direct comparison, by an example.

We have studied two examples of amorphous-coated wires: a 1-D model, and silica coated Si nanowires.
In the examples studied, specific physical phenomena for coated systems have arosen: 

1) {\it Ballistic-diffusive transition:} 
We explicitly showed the evolution of the phonon transmission
in the transitional regime, before $\lambda(\omega,L)$ asymptotically approaches its limiting value. 
In the limit of long coated length the
transmission attains a $1/L$ behavior. 
(In contrast, for rough-boundary uncoated wires, an exponential decrease with length takes place \cite{Santamore}.)

2) {\it Low temperature conductance:} 
near $\omega=0$, the transmission is found to decrease with respect
to its $\omega=0$ value, as the frequency is increased. As a result, amorphous-coated wires
display a low temperature dip in the plot of the thermal conductance divided by temperature, having certain
similarities with the dip reported for rough uncoated wires \cite{Santamore}.

3) {\it Coating-wire interface structure effects:} 
a new phenomenon is found, which is specific of 
coated wires: as the interatomic bonds between wire and coating atoms are made stronger, starting from zero,
the transmission of a long coated wire (or equivalently, $\lambda(L,\omega)$) decreases, 
but only up to a certain limit; once the bond strength reaches a certain threshold,
the transmission {\it increases} again as the bond is made stronger. When the bond strength tends to infinity
the relaxation length asymptotically saturates to a finite value. 
An analogous behavior was found to take place as a function of the number of interface links. The structure
of the interface was shown to have an important effect on the relaxation length.

It is hoped that the method and results presented here will help to estimulate further atomistic research 
in the field of phonon transmission through nanowires, as well as experimental research on amorphous-coated
nanowires.

\begin{acknowledgments}
We acknowledge Toshishige Yamada for valuable comments on the manuscript, and also A.~Balandin and P.~Schelling
for useful discussions.
\end{acknowledgments}

\appendix*
\section{The role of anharmonicity}
In one dimension, there is no three phonon anharmonic scattering, since
it is not possible to have a process in which all three phonons belong to the same 
polarization branch \cite{Ziman}. 
Therefore, anharmonicity is not an issue in the first example, given in section~\ref{sec:1dwire}.

The role of anharmonicity in the thermal conductivity of Si nanowires has been studied in
ref.~\onlinecite{Mingo}. For the wires studied in section~\ref{sec:sinanow}, its influence is 
negligible. An estimation can be made using the Mathiessen rule. The total relaxation length,
expressed in terms of the boundary and anharmonic relaxation lengths, is: 
\begin{eqnarray}
\lambda=1/(\lambda_b^{-1} + \lambda_a^{-1}). 
\label{eq:appendix}
\end{eqnarray}
Now, $\lambda_b\sim D$, where $D$ is the system's
thickness (around 2 nm, for the wires considered here). Also, 
$\lambda_a \sim c/(1.73\times 10^{-19}{s\over K} T \omega^2 e^{-137.3K/T})$, where
$c$ is the Si speed of sound \cite{Mingo}. 
The role of anharmonicity
is largest for the highest frequencies. But even for the highest frequency available in Si, $\sim 95$~THz,
the anharmonic relaxation length at room temperature is still $\lambda_a \sim 20$~nm. This, 
in the case of the nanowires of section~\ref{sec:sinanow}, using eqn.~(\ref{eq:appendix}) and the
estimate just given for $\lambda_b$,
affects the transmission by less than 10\%. (Lower frequencies are much less affected.) The
influence on the thermal conductivity is much smaller than this, since high modes contribute significantly
less than the lower energy modes. Experimentally it has been observed that even for nanowires
37 nm thick, the effect of anharmonicity on the thermal conductivity is minimal compared to
that of boundary scattering, up to 300K. This is evident from the fact that the
thermal conductivity stays constant with T near room temperature, rather than decrease 
(see refs.~\onlinecite{Li,Mingo}).


\begin{thebibliography}{1-100} 
\bibitem{Lieber} Yi Cui, Lincoln J. Lauhon, Mark S. Gudiksen, Jianfang Wang, and Charles M. Lieber,
Appl. Phys. Lett. {\bf78}, 2214 (2001).
\bibitem{Schwab} K. Schwab, E. A. Henriksen, J. M. Worlock, and M. L. Roukes,
Nature, {\bf 404}, 974 (2000).
\bibitem{Fon} W. Fon, K. C. Schwab, J. M. Worlock, and M. L. Roukes, Phys. Rev. B {\bf 66}, 045302 (2002).
\bibitem{Kim} P. Kim, L. Shi, A. Majumdar, and P. L. McEuen, Phys. Rev. Lett. {\bf 87}, 215502 (2001) 
\bibitem{Balandin} J. Zou and A. Balandin, J. Appl. Phys. {\bf89}, 2932 (2001); 
A. Balandin and K. L. Wang, Phys. Rev. B {\bf58}, 1544 (1998).
\bibitem{Volz} S. G. Volz and G. Chen, Appl. Phys. Lett. {\bf75}, 2056 (1999).
\bibitem{Schelling} P. K. Schelling, S. R. Phillpot, P. Keblinski, Phys. Rev. B {\bf65}, 144306 (2002).
\bibitem{Kambili} A. Kambili, G. Fagas, V. I. Fal'ko and C. J. Lambert, Phys. Rev. B 
{\bf60}, 15593 (1999).
\bibitem{Ciraci} A. Ozpineci and S. Ciraci, Phys. Rev. B {\bf63}, 125415 (2001);
A. Buldum, S. Ciraci, and C. Y. Fong, J. Phys.: Condens. Matter {\bf 12}, 3349 (2000);
S. Ciraci, A. Buldum, and I. P. Batra, ibid. {\bf 13}, R537 (2001).
\bibitem{Rego} L. G. C. Rego and G. Kirczenow, Phys. Rev. Lett. {\bf 81}, 232 (1998);
Ibid. {\bf 81}, 5037 (1998).
\bibitem{Santamore} D. H. Santamore and M. C. Cross, Phys. Rev. B {\bf 63}, 184306 (2001);
D. H. Santamore and M. C. Cross, Phys. Rev. B {\bf 66}, 144302 (2002);
D. H. Santamore and M. C. Cross, Phys. Rev. Lett. {\bf 87}, 115502 (2001).
\bibitem{Angelescu} D. E. Angelescu, M. C. Cross, and M. L. Roukes,
Superlattices and Microstruct., {\bf 23}, 673 (1998).
\bibitem{Blencowe} M. P. Blencowe, Phys. Rev. B {\bf 59}, 4992 (1999).
\bibitem{Leitner} D. M. Leitner and P. G. Wolynes, Phys. Rev. E {\bf 61}, 2902 (2000).
\bibitem{Patton} K. R. Patton and M. R. Geller, Phys. Rev. B {\bf 64}, 1555320 (2001).
\bibitem{Ma},D. D. D. Ma, C. S. Lee, F. C. K. Au, S. Y. Tong, and S. T. Lee,
Science {\bf 299}, 1874 (2003).
\bibitem{Glavin} B. A. Glavin, Phys. Rev. Lett. {\bf86}, 4318 (2001).
\bibitem{Economou} E. N. Economou, Green's Functions in Quantum Physics, Springer-Verlag,
Berlin (1983).
\bibitem{Klemens} P. G. Klemens, Solid State Physics, edited by F. Seitz and D. Turnbull; 
P. G. Klemens, Proc. R. Soc. London, Ser. A {\bf208}, 108 (1951).
(Academic, New York, 1958), Vol. 7, p.1.
\bibitem{Verges} L. Mart\'in Moreno and J. A. Verg\'es, Phys. Rev. B {\bf 42}, 7193 (1990).
\bibitem{Laughlin} R. B. Laughlin and J. D. Joannopoulos, Phys. Rev. B {\bf16}, 2942 (1977).
\bibitem{Landauer} Y. Imry and R. Landauer, Rev. Mod. Phys. {\bf71}, S306 (1999).
\bibitem{Jones} W. Jones and N. H. March, {\it Theoretical Solid State Physics}, Dover (1985).
\bibitem{Mingo01} N. Mingo, Liu Yang and Jie Han, J. Phys. Chem. B, 105(45), 11142 (2001).
\bibitem{Goldberg} E. C. Goldberg, A. Mart\'in-Rodero, R. Monreal and F. Flores, Phys. Rev. B {\bf 39},
5684 (1989).
\bibitem{Guinea} F. Guinea, C. Tejedor, F. Flores and E. Louis,
Phys. Rev. B 28(1983)4397.
\bibitem{Callaway} J. Callaway, Phys. Rev. {\bf113}, 1046 (1959).
\bibitem{Holland} M. G. Holland, Phys. Rev. {\bf132}, 2461 (1963).
\bibitem{Asen-Palmer}M. Asen-Palmer, K. Bartkowski, E. Gmelin, M. Cardona, A. P. Zhernov,
A. V. Inyushkin, A. Taldenkov, V. I. Ozhogin, K. M. Itoh and E. E. Haller,
Phys. Rev. B {\bf56}, 9431 (1997).
\bibitem{Beenaker}C. W. J. Beenakker, Rev. Mod. Phys. {\bf69}, 731 (1997).
\bibitem{deJong} M. J. M. de Jong, Phys. Rev. B {\bf 49}, 7778 (1994).
\bibitem{Ziman} J. M. Ziman, in {\it Electrons and Phonons}, Oxford (1963).
\bibitem{Harrison} W. A. Harrison, Electronic Structure and the Proporties of Solids,
Dover, New York (1989).
\bibitem{Weber} W. Weber, Phys. Rev. B {\bf15}, 4789 (1977).
\bibitem{Siband} G. Nilsson and G. Nelin, Phys. Rev. B 6, 3777-3786 (1972).
\bibitem{Silica} C.-K. Loong, J. European Ceramic Society {\bf19}, 2241 (1999).
\bibitem{branches} N. Nishiguchi, Y. Ando and M. N. Wybourne, J. Phys. Condens. Matter 
{\bf9}, 5751 (1997)
\bibitem{Mingo} N. Mingo, Phys. Rev. B, in press. Also, cond-mat/0308587.
\bibitem{Li} D. Li, Y. Wu, P. Kim, L. Shi, P. Yang, and A. Majumdar, Appl. Phys. Lett., in press.
\end{thebibliography}
\end{document}